\documentclass[aps,pra,nofootinbib,preprint,amsmath,amssymb,floatfix]{revtex4-2}
\usepackage{color}
\usepackage{graphicx}

\begin{document}
\count\footins = 1000

\newcommand{\tc}{\textcolor}
\newcommand{\g}{blue}
\title{Below the Schwinger critical magnetic field value, quantum vacuum and gamma-ray bursts delay}
\author{ Iver H. Brevik$^1$, Moshe M.  Chaichian$^2$ and Anca Tureanu$^2$  }      
\affiliation{$^1$Department of Energy and Process Engineering, Norwegian University of Science and Technology, N-7491 Trondheim, Norway\\
$^2$Department of Physics, University of Helsinki, and Helsinki Institute of Physics,  P. O. Box 64, FI-00014 Helsinki, Finland}


\begin{abstract}
A magnetic field above the Schwinger critical value $B_{\rm crit} = 10^9$ Tesla is much higher than any magnetic field known by now in the interstellar bulk except in the vicinity of observed magnetars having magnetic fields between $10^9$ and $10^{11}~$Tesla. Above the critical magnetic field limit, calculated by Schwinger in the lowest order perturbation in quantum electrodynamics (QED), one reaches the threshold for electron-positron pair creation {(through the intermediate electric field, as known also from standard electrodynamics)}, which has interesting consequences. Therefore, finding out whether one could encounter some consequences of interest also for the values of the magnetic field below the Schwinger critical point, we invoke the next higher-order effect in QED, which is emerging from  the {\it Quantum Vacuum Effect}. The latter is equivalent to the use of the Euler-Heisenberg effective theory in nonlinear electrodynamics, where the Lagrangian has now a term with a higher power, $B^4$. In this case, in the region $B<B_{\rm crit}$, we show that interesting effects appear, among them the Cherenkov radiation and the reduction in the speed of light. The latter effects appear because of the quantum vacuum mimicking a medium. We also present quantitative arguments for such a close analogy. As a rough estimate, we show that the time delay $\tau$ of gamma-ray bursts (GRB) having traveled through the entire cosmological distances in an average strong magnetic field such as  $10^6~$Tesla, reaches an experimentally considerable value of $\tau = 2.4$ hours. Of course in the vicinity of magnetars, the magnetic field is much stronger, of the order of $10^9-10^{11}$ Tesla. However, in this case the linear scale of GRB trajectory through such regions would be much smaller. For the latter, we also give a corresponding estimate for the number of the magnetars along the trajectory and also for the delay. Finally, we shall dwell on the recently raised issue in the literature, namely the Lorentz invariance violation (LIV).
\end{abstract}
\maketitle

\section{Introduction}
\label{secintro}

As is known, when an electromagnetic field in vacuum is exposed to an external strong static magnetic field ${\bf B}_0$, the equations of motion derived from the generalized high-intensity action integral become formally analogous to the conventional Maxwell equations in a continuous medium \cite{berestetskii82}.  Then it becomes natural to also introduce nondispersive dielectric constants $\varepsilon_{ik}$ and $\mu_{ik}$, which in turn lead to the refractive index components,  greater than unity.

The Schwinger critical limit $B_{\rm crit}$ \cite{schwinger51} occurs when the magnetic field $B_0$ becomes so strong that electron-positron pair production occurs. The value $B_{\rm crit}= 10^9~$Tesla is far beyond what is achievable on Earth, while it is known to be surpassed in the vicinity of astrophysical objects such as magnetars, first announced in 2008 \cite{ESO08}.  It is of interest to investigate also the case for the more accessible lower magnetic fields $B<B_{\rm crit}$, and to explore the implications of this kind of medium-like vacuum, appropriately called the {\it quantum vacuum effect}.  

From a fundamental viewpoint, one notices that a constant magnetic field alone will
not be able to produce electron-positron pairs. One needs an electric field as an intermediate
factor. This is essentially the same kind of behaviour as encountered in standard electrodynamics -- an electric field varying in space is needed so as to get a magnetic field developing in time, according to Maxwell's equations.

As mentioned above, the effect is related to the Euler-Heisenberg theory of nonlinear electrodynamics \cite{heisenberg36}, derived from the higher-order corrections in QED.  An extensive study of the effect from a quantum perspective was given by Adler \cite{adler71}. The propagation of photons in an external field and photon splitting were considered in Refs.~\cite{bialynicki70} and \cite{brezin71}. Reference \cite{dittrich98} considered, from a geometric optics approximation, various nontrivial QED vacua. From an inflationary viewpoint, the possibility of generation of the origin of primordial magnetic field was discussed in Ref.~\cite{demozzi09}. As well, the study of the Quantum Magnetic Collapse in a strong magnetic field of the order of the critical magnetic fields has been performed in \cite{Chaichian-Montonen}, with possible implications of the results for astroparticle physics and cosmology. Also there exist some works on the effect of homogeneous magnetic fields on the refractive index \cite{tsai74,tsai75}. In the latter works two different cases were considered: one in which high-energy photons propagate in fields weak compared with the critical field, and one where low-energy photons propagate in fields with arbitrary intensity.

  In connection with the nonlinear electrodynamics effects,  in \cite{12'} and \cite{12''}, respectively the photon polarization and the vacuum birefringence have been considered in detail, while in  \cite{ferrer84} the  polarization selection rules that take place in the photon splitting in a medium (i.e. in the presence of a chemical potential) were discussed.

There also exist studies of the supercritical limit i.e., when $B_0>B_{\rm crit}$ \cite{shabad04,16'}.  An earlier work of the same group of authors had studied the case when there were both a constant electric and a constant magnetic field \cite{batalin71}.

In the following we shall first give an overview of the electromagnetic properties of this kind of magnetic-generated medium in the case of a subcritical magnetic field $B_0 < B_{\rm{crit}}$, and will thereafter focus on some new aspects not considered previously in the literature.  Specifically:

\noindent {{\it i)} Considering that photons propagating in a medium are slowed down to the velocity $c/n$ (here assuming that the medium is isotropic), one may ask: how large must $B_0$ be in order to make the effect observable? How does the photon velocity reduction, if calculated on the basis of the mean magnetic field in intergalactic space, compare with the dispersive slowing down of the velocity of photons in the  electron plasma, as considered in \cite{chaichian20,brevik21}?

\noindent {\it ii)} As a rough estimate,  we can take the average magnetic field to be  $10^6~$Tesla, which although seems to be
high, nevertheless is still well below the Schwinger critical limit, extending over a large cosmic
distance. In such a case, the delay of GRB, $\tau$, comes out to be large, estimated of the order of $\tau =$ 2.4 hours. Therefore, when two species of particles, namely photons and neutrinos, are traveling through the interstellar distances, photons arrive later by a time delay $\tau$ compared with neutrinos, which have no interaction with the magnetic field. However, as mentioned in the Abstract, in the vicinity of magnetars magnetic field is much stronger, of the order of $10^9-10^{11}$ Tesla. With such values of the magnetic fields we also give an estimate  for the delay, considering a realistic case with the number of magnetars along the trajectory passed by the GRB in the Subsection B of Section III. Actually several earlier  observations  such as  \cite{A,B} and references therein led
to the idea of a Lorentz invariance violation (LIV), assuming that the two different species of neutrino and GRB
were  emitted from the same source and at the same time. The important point, however, is that in those observations there was no assurance of any kind in their accuracy of measurements that the two species were emitted from the same source, not speaking of their simultaneous emission. In addition to all that,  the measured delays were quite inaccurate. Still those observations gave boost to a possible LIV. Concerning the claimed  observation  of LIV  based on  different delays depending on the energy spectrum of emitted GRB, and the implication of the present work, we present a full  detail in the Section \ref{sec:final}, where we give arguments that   there has  been   no evidence   for the LIV extracted from the GRB spectrum.


\noindent { \it iii)}  Since the theoretical value of $n$ is greater than one, the possibility of a Cherenkov radiation   becomes of immediate interest. Under terrestrial conditions, such an effect appears to be far too small to be observed. But what can be said about the specific case of {\it atmospheric
constellation}, and whether there is any situation, where such an effect as the Cherenkov radiation can be detected? Apparently, there are only a few works in the literature discussing such a  possibility (see the recent paper  \cite{macleod19} and references therein).



In the next section, we give an overview of the general theory describing in the subcritical case, where the magnetic field is lower than the critical value.  In this work we use the Gaussian units, mostly with $\hbar = c = 1$, and follow the notation of  Ref~\cite{berestetskii82}. Finally, we present several new results and discuss the relation to their observability.

\section{Basics }\label{sec:basics}

Let the total  magnetic field in a homogeneous region of outer space be denoted   by $\bf B$. It is composed of the strong static field ${\bf B}_0$ and a weak time-varying part ${\bf B}'$ associated with the traveling electromagnetic waves. No dielectric media are assumed present. The Schwinger critical limit is
\begin{equation}
B_{\rm crit} = \frac{m^2c^3}{|e|\hbar}= 4.41\times 10^9~{\rm Tesla} =  4.41\times 10^{13}~{\rm Gauss}. \label{1}
\end{equation}
 In this formula, $m$ is the electron mass and $e$ is the electron charge. We re-emphasize that, while the critical field is far too strong to be producible on Earth, it is significantly exceeded in magnetars ($10^9 - 10^{11}~$Tesla)\footnote{ If one would consider axion electrodynamics (cf., for instance, Ref.~\cite{sikivie21}), the form of $B_{\rm crit}$ would be different from Eq.~(\ref{1}) and it could happen that the value of $B_{\rm crit}$ would drastically change and even become much reduced. The case with axion electrodynamics is of interest to be investigated further. The latter can be of interest also due to an excess in the amount of axions around the magnetars as recently argued in \cite{19'}.}.

It is in principle straightforward to consider a strong electric field ${\bf E}$ in the same way. The critical value becomes then
\begin{equation}
E_{\rm crit}= \frac{m^2c^3}{|e|\hbar}= 1,32\times 10^{18}~{\rm Volt/m}.
\end{equation}
We will be concerned with low-order nonlinear theory, meaning that the fields are obtained from first order expansions in the parameter $(B_0/B_{\rm crit})^2$.
 We start from the perturbed Lagrangian density $L'$, derived from the original (unperturbed) Maxwell Lagrangian density $L$.  In the general case \cite{berestetskii82},
\begin{equation}
L' = \frac{e^4}{360 \pi^2m^4}[ (E^2-H^2)^2+7({\bf E\cdot H})^2].
\end{equation}
In  the following we shall assume that there is no external electric field, i.e. ${\bf E}=0.$ Thus,
\begin{equation}
L'= \frac{e^4}{360 \pi^2m^4}H^4,
\end{equation}
which enables us to calculate the magnetization $\bf M$ in the  medium by using the general formula ${\bf M}=\partial L'/\partial {\bf H}$,
\begin{equation}
{M}= \frac{e^4}{90\pi^2m^4}H^3. \label{magnetization}
\end{equation}
Now one has ${\bf B}=\mu {\bf H}$ where  $\mu$ is the permeability. Introducing   the susceptibility  $\chi$   so that ${\bf M}=\chi {\bf H}$, we have in Gaussian units
\begin{equation}
\chi= \frac{\mu-1}{4\pi},
\end{equation}
leading to
\begin{equation}
\mu = 1+ \frac{2e^4}{45\pi m^4}H^2. \label{permeability}
\end{equation}
{Along with the permeability $\mu$, the general formalism introduces also a constant permittivity $\varepsilon$ in the same way; cf. Ref. \cite{berestetskii82}. These two material parameters have thus the same
fundamental basis. (The details of the derivation are given below, see formulas \eqref{B-E fields}-\eqref{7}.)}
The corresponding expression for the permittivity $\varepsilon$ is
\begin{equation}
\varepsilon = 1+ \frac{5e^4}{45\pi m^4}H^2.
\end{equation}
Thus, the refractive index $n$ becomes
\begin{equation}
n = \sqrt{\varepsilon \mu}=  1+\frac{7e^4}{90\pi m^4}H^2.
\end{equation}
Note: this simple calculation rests upon the assumption that it is physically meaningful to consider the medium as isotropic, possessing scalar expressions for $\varepsilon$ and $ \mu$. This is not evident beforehand, but will be justified by a more detailed consideration below. The main aim of the calculation is to show that the two constants of the matter, $\mu$ and $\varepsilon$ are quadratically dependent on the applied magnetic field\footnote{There is an extra $\pi$ in the denominators of the above formulas, compared with those given in Ref.~\cite{berestetskii82}.}.

We now consider the physical situation in more detail, focusing on the weak fields  ${\bf E', B'}$  associated with propagating waves in the strong magnetic field background. With the decomposition
\begin{equation}\label{B-E fields}
  {\bf B} = {\bf B}_0+{\bf B'}, \quad {\bf E} = {\bf E'},
\end{equation}
assuming a common time factor $e^{-i\omega t}$,
we then obtain from Maxwell's  equations
\begin{equation}
{\bf k\times H'} = -\omega {\bf D'}, \quad {\bf k\times E'}= \omega {\bf B'},
\end{equation}
\begin{equation}
{\bf k \cdot B'}=0, \quad {\bf k\cdot D'}=0,
\end{equation}
corresponding to the constitutive relations for an anisotropic medium,
\begin{equation}
D_i'= \varepsilon_{ik}E_k',  \quad  B_i' = \mu_{ik}H_k'.
\end{equation}
Explicitly, the matter's constants can be expressed as \cite{berestetskii82}
\begin{equation}
\varepsilon_{ik}= \delta_{ik}+\frac{2e^4}{45 m^4}B_0^2(-\delta_{ik}+\frac{7}{2}b_ib_k), \label{4}
\end{equation}
\begin{equation}
\mu_{ik}= \delta_{ik}+\frac{2e^4}{45m^4}B_0^2(\delta_{ik}+2b_ib_k), \label{5}
\end{equation}
where ${\bf b}={\bf B}_0/B_0$ is the unit vector in the direction of the applied field. We can express
these equations also in terms of the critical field,
\begin{equation}
\varepsilon_{ik}= \delta_{ik}+\frac{2e^2}{45}\frac{B_0^2}{B_{\rm crit}^2} (-\delta_{ik}+\frac{7}{2}b_ib_k ), \label{6}
\end{equation}
\begin{equation}
\mu_{ik}= \delta_{ik}+\frac{2e^2}{45}\frac{B_0^2}{B_{\rm crit}^2}  \left( \delta_{ik}+2b_i b_k \right). \label{7}
\end{equation}
In Gaussian  units,  $e^2=  1/137$\footnote{There is a debate in the literature (see \cite{Jeremy}) concerning the numerical factor in the second term within the brackets of Eq. \eqref{5}, implying a numerical change compared with the corresponding values given in \cite{berestetskii82}. However, such minor numerical change in no way would alter the main results presented in our work. We thank Yuri Obukhov for bringing this point to our attention.}.

As mentioned above, we limit ourselves to the case of weak nonlinearity, meaning that the  condition for the use of Eqs.~(\ref{6}) and  (\ref{7}) is $B_0/B_{\rm crit} \ll 1$.    The effective magnetic expansion parameter is \cite{berestetskii82}
\begin{equation}
eH/m^2 \ll 1, \label{8}
\end{equation}
with $H$ the total magnetic field. In practice, $H$ can here be replaced by $B_0$.

We may distinguish between two cases of linear polarization of a propagating plane wave, taking into account the special nature of the ${\bf bk}$ plane. We shall however focus mainly on   the perpendicular case, designated by a subscript $\perp$, where ${\bf B'}$ is perpendicular to the mentioned plane. For simplicity we take ${\bf B}_0$ in the $x$ direction, and the wave vector $\bf k$ in the $z$ direction. For later convenience we define  the symbol $d$  as
\begin{equation}
d=     \frac{7e^4}{90m^4}B_0^2 =      \frac{7e^2}{90}\frac{B_0^2}{B_{\rm crit}^2}. \label{9}
\end{equation}
Therefore, from Eqs. \eqref{6} and \eqref{7}, we obtain
\begin{equation}
\varepsilon_{xx}= 1+\frac{10d}{7}, \quad \varepsilon_{yy} = \varepsilon_{zz}= 1-\frac{4d}{7}, \label{10}
\end{equation}
\begin{equation}
\mu_{xx}= 1+\frac{12d}{7}, \quad \mu_{yy}=\mu_{zz}=1+ \frac{4d}{7}, \label{11}
\end{equation}
the nondiagonal components $\varepsilon_{ik}$ and $\mu_{ik}$ being equal to zero.

Here the most significant constitutive relation is
\begin{equation}
B_y' = \mu_{yy}H_y' = \left( 1-\frac{4d}{7}\right)H_y',
\end{equation}
since it corresponds to the real wave for  $B_y'$. We see that  $\mu_{yy}$ is equal to the permeability $\mu$ in the isotropic model above, Eq.~(\ref{permeability}), to the approximation considered. This justifies the introduction of the isotropic model. Once $\mu_{yy}$ is fixed, the other components $\varepsilon_{ik}$ and $\mu_{ik}$ follow from Maxwell's equations.

With $k= n_\perp \omega$, $n_\perp$ being the refractive index for perpendicular polarization we thus have, with $n_\perp = \sqrt{\varepsilon \mu}$,
\begin{equation}
n_\perp = 1+d =   1+ \frac{7e^4}{90m^4}{B_0^2} =     1+ \frac{7e^2}{90}\frac{B_0^2}{B_{\rm crit}^2}. \label{14}
\end{equation}

In the same way one can analyze the case of parallel polarization, subscript $\parallel$ as mentioned. This case leads to a refractive index $n_\parallel$, which is smaller than $n_\perp$,
and will not be further considered here.

We re-emphasize that the present dielectric  theory, coming from the quantum vacuum, is nondispersive, i.e there is no frequency
dependence in the matter’s constants and thus in the refraction index too.


\section{Applications}

We shall now illustrate the use of the above formalism, under various circumstances.

\subsection{Reduction of the photon velocity in the medium (analogous to the works in \cite{chaichian20,brevik21}). The weak magnetic field case}

As is the case in all dielectric media, the propagation velocity is lower than in a vacuum. We exploit our earlier definition of the parameter $d$ in Eq.~(\ref{9}) to write the velocity in the simple form
\begin{equation}
v= c/n = c(1-d).  \label{15}
\end{equation}
Numerically, from Eq.~(\ref{9}) we get, inserting  $B_{\rm crit}$ from Eq.~(\ref{1}),
\begin{equation}
d= 2.9\times 10^{-31} \times B_0^2, \label{16}
\end{equation}
where $B_0$ is in gauss (G) units. Taking $B_0= 100~\mu G$ to be representative for intergalactic space,  we obtain
\begin{equation}
v= c(1-2.9\times 10^{-39}). \label{17}
\end{equation}
This illustrates how extremely small is the reduction of the light velocity under these conditions.  However, if an appreciable light reduction is to be achieved, one has to focus on  local astrophysical regions where $B_0$ is very strong. It is to be noted that even with a field of $B_0= 10^{10}~$G, very strong according to normal standards, we are still far below the critical limit (\ref{1}).

Related to this point, it is of interest also to compare this case with the results one gets by looking at how much photons are slowed down because of the electron plasma in the interstellar space, i.e.  without having a magnetic field.   This means to compare with the velocity reduction of gamma-rays due to the conventional dispersive theory.  We take the electrons with mass $m$ and their mean density in the interstellar space to be $N$.

We start with  the dispersive relation for photons in an electron(-positron) plasma,
\begin{equation}
\omega^2 = k^2+ \omega_p^2, \label{18}
\end{equation}
where $\omega_p$ is the plasma frequency determined by
\begin{equation}
\omega_p^2 = \frac{4\pi Ne^2}{m}.  \label{19}
\end{equation}
The system is thus dispersive, and we consider the group velocity
\begin{equation}
v_g = \frac{d\omega}{dk}= c\left( 1-\frac{\omega_p^2}{2\omega^2}\right). \label{19}
\end{equation}
Here  we insert the electron mass $m=0.511~$MeV, and, similarly as in Ref.~\cite{brevik21}, focus attention on  galaxy filaments at around $z=0.1$, for which the electron density is about $N= 4\times 10^{-4}~$cm$^{-3}$. In view of the conversion factor 1 cm $= 5.068\times 10^{13}~$GeV$^{-1}$ this means $N= 3.1\times 10^{-45}~$GeV$^3$ so that $\omega_p^2 = 5.5\times 10^{-43}~$GeV$^2$ = $1.2\times 10^6~$s$^{-2}$. Thus, for a gamma-ray with energy $E= 100~$GeV\footnote{In order not to violate the parameter regime where the Heisenberg--Euler Lagrangian as the leading-order in the perturbative expansion can be reliably applied, besides the condition \eqref{8} $eB/m^2 \ll 1$ , there is still another criterion \cite{12''}, namely the condition
\begin{equation}\frac{e B \omega}{m^3} \ll 1.\end{equation}     
With a value of $E= 100~$GeV  for the energy of GRB, one gets  $e B \omega/m^3=44.4$, which is clearly not smaller than unity. However, with lower energies,  such as $E= 1~$GeV, one can trust the leading term in the perturbative expansion as performed in the present work, and the qualitative results coming from them.
As a result, for the GRB energy $E= 1~$GeV,  the  delay times   calculated  for $E= 100~$GeV, i.e. Eq. \eqref{34'} and Eq. \eqref{eq35}, will be reduced by the corresponding factor of 100.

Since the main result of the work is to show that there is no LIV derived from GRB  delays,  contrary to some opposite  claims in the literature, the nondispersive character, emerging as a by-product of calculations and  mentioned in the work, is not of importance for our arguments, which  are  based entirely on the emission of different frequencies of light from their source occurring NOT simultaneously.} we obtain
\begin{equation}
v_g= c(1-3\times 10^{-47}),
\end{equation}
which shows an even smaller velocity reduction  than the previous  expression (\ref{17}).

\subsection{ Reduction of the photon velocity in the medium in the presence of a strong magnetic field.
The delay of photon detection time}

Let us assume now an extreme case when the GRB photons travel over large intergalactic distances, experiencing strong magnetic fields all the time, before they each us, i.e. the detectors on the Earth. For definiteness, we choose an average value for the magnetic field as mentioned above, i.e.
\begin{equation}
B_0 = 10^{10}~\rm{G},
\end{equation}
and make a crucial assumption  that this magnetic  field is constant over the  {\it entire} photon trajectory. At the end of this Subsection we present a realistic case, where the trajectory of photon is considered to pass only  through  the vicinity of the magnetars, with an estimate on their number in the interstellar space. 
As compared with the value $100~\mu$G used in the previous subsection we are thus increasing $B_0$ by a factor of $10^{14}$. We obtain now
\begin{equation}
v=c( 1-2.9\times 10^{-11}).
\end{equation}
This number is actually of considerable physical interest: what is the  {\it time delay} of a gamma-ray burst having traveled a large intergalactic distance in such a  strong field before it reaches us? Denoting the effective distance by $D$ we obtain for the delay time $\tau$, omitting for simplicity the redshift  correction,
\begin{equation}\label{(33)}
\tau = \frac{D\times d}{c(1-d)} \approx \frac{D\times d}{c}.
\end{equation}
As in  Ref.~\cite{brevik21} we choose for $D$ the large value $3\ \rm{Gpc}=9\times 10^{25}~$m, whereby
\begin{equation}\label{34'}
\tau = 8.7\times 10^6~{\rm s}= 2.4~{\rm hours}.
\end{equation}

As mentioned  before, at the  point ii) above, the delay time $\tau$  in Eq. \eqref{34'} is of considerable importance to explain  the delay of the GRB compared  with the neutrino bursts, without invoking the LIV.

The assumption made above on the choice of $B_0= 10^6\ \mbox{Tesla}$ as an average  cosmological magnetic field in the {\it entire} interstellar space, seems to be too strong. However,  there are regions where still stronger magnetic fields, of the order of $10^9$ to $10^{11}$ Tesla exist in the vicinity of magnetars, but the linear scale of such regions would be much smaller, around  magnetars not exceeding some $10^9$ cm or so. Assuming now that the mean separation between galaxies is about $1\ \mbox{Mpc}=3\times10^{24}\ \mbox{cm}$, then there could be  about $10^4$ galaxies along the line of sight, passing the entire interstellar distance of $3\ \mbox{Gpc}=10^{28}\ \mbox{cm}$.

In such a  case, instead of the time delay as $\tau = 8.7 \times 10^6\ \mbox{s} = 2.4\ \mbox{hours}$ according to Eq. \eqref{34'}, we shall have a time delay to be anything  between the values  such as
\begin{equation}\label{eq35}\tau = 8.7\times (10^2 - 10^{18})\ \mbox{s}
\end{equation}
according to the Eqs. \eqref{16} and \eqref{(33)}. In the estimate \eqref{eq35} we have also  taken  into account that within the sphere around  each magnetar, in average the magentic field value is reduced by a factor of $10^2$ from  $B_0= 10^9 -10^{11}\ \mbox{Tesla}$. We are aware of the fact that  the  range of values given  in \eqref{eq35}  as compared with the neutrino bursts, are far from being realistic due to the unknown  number of magnetars in the galactic space and as well the value of magnetic fields around them\footnote{For a celestial experiment, one  might  think of a Fizeau-type setup, in which the whole system is placed in a constant strong magnetic field to observe the reduction  in the light velocity.}. But the  observations in comparing the neutrino with  GRB bursts are also  not accurate and  conclusive by any standard all.

Let us also mention that the delay of neutrinos traveling through the entire galactic space  is fully negligible, as  shown in \cite{chaichian20,brevik21} taking into account the masses of the three species and their oscillations.

\subsection{Remarks on the  Cherenkov effect}

The fact that $n>1$ in the nonlinear theory makes it natural, as mentioned above, to investigate the possibility of observing a Cherenkov effect although the smallness of the medium susceptibility necessarily imposes strong restrictions on the particle velocity. To our knowledge there are only a few earlier investigations along these lines, {although we
may mention Refs. ~\cite{macleod19, Lee}, and also the older review article Ref.~\cite{Erber}.} It becomes now natural to consider protons instead of electrons as the fundamental particles. One may note that with the proton mass $m_p= 938~$MeV the  critical magnetic field, Eq.~(\ref{1}), becomes enhanced by a factor $(m_p/m)^2$.

It is obvious that we have to look for extreme events in the intergalactic space.  Cosmic radiation in the form of protons has been observed up to about $3\times 10^ {20}~$eV. For definiteness, let us consider the "window" where the proton energy is
\begin{equation}
E > 10^{15}~{\rm eV},
\end{equation}
and investigate under which conditions Cherenkov radiation in this energy interval can occur. A proton  with energy $E= 10^{15}~$eV  in rectilinear uniform motion  has a velocity $\beta$ equal to
\begin{equation}
\beta = c\left[ 1-\frac{1}{2}\left( \frac{m_p}{E}\right)^2\right] = c(1-4.4\times 10^{-13}). \label{beta}
\end{equation}
As before, we consider the case of perpendicular polarization, where the component $B_y'$ propagates in the $z$ direction in the nonlinear  medium created by the $x$ directed magnetic field $B_0$. The threshold for the Cherenkov effect is
\begin{equation}
n_\perp \beta =1.
\end{equation}
Thus the threshold corresponds to the perpendicular refractive index
\begin{equation}
n_\perp = \frac{1}{\beta} = 1+ 4.4\times 10^{-13},
\end{equation}
where use is made of Eq.~(\ref{beta}). {We compare this with the equation \eqref{14}, i.e. with
\begin{equation}\label{eq_40}
n_\perp = 1+ \frac{7e^4}{90 m^4}B_0^2,
\end{equation}
where the mass of the electrons $m$ enters, because the medium, as before, is the quantum electrodynamics vacuum, which gives the dominant contribution as compared to the contributions coming from the other constituents present in the quantum vacuum. Then, from Eq. \eqref{eq_40}, we derive for the magnetic field $B_0$ the value:
\begin{equation}
B_0= 4.5\times 10^{-14}~{\rm GeV}^2.
\end{equation}}
Using the conversion factors 1 GeV= $1.602\times 10^{-3}~$erg, 1 cm = $5.068\times 10^{13}~$GeV$^{-1}~$, we have the following relationship between the energy densities in the natural and the Gaussian system of units,
\begin{equation}
{\rm GeV}^4= 2.08\times 10^{38}~{\rm \frac{erg}{cm^3}}.
\end{equation}
Since the magnetic energy density in Gaussian units is $B_0^2/(8\pi)$, we obtain the magnetic threshold value to be
\begin{equation}
B_0= 6.5\times 10^{5}~{\rm G}.
\end{equation}
This appears to be roughly as we might expect: this value is less than the critical field even for electrons (cf. Eq.~(\ref{1})), and is much less than the magnetic field observed around the magnetars. So protons with energies in the mentioned energy "window" will in principle be able to initiate Cherenkov radiation if they pass through regions with so strong magnetic fields.

Finally, we give the Tamm-Frank seminal formula for  the emitted Cherenkov radiation from a particle with charge $e$ moving with velocity $v$ along the $x$ axis, in Gaussian units,
\begin{equation}
\frac{dE_{ch}}{dx} = \frac{e^2}{4\pi}\int_{\omega > c/n(\omega)} \mu(\omega)\omega \left( 1-\frac{c^2}{v^2n^2(\omega)} \right)d\omega,
\end{equation}
where $\mu(\omega)$ is the magnetic permeability of the medium, which is now given by the Eqs. \eqref{9} and \eqref{11}. This formula can be used for a possible experimental detection of the Cherenkov radiation.

\section{Discussion and Conclusions}\label{sec:final}

The quantum vacuum mimicking a medium carries over to all the characteristic properties of conventional electrodynamics of media, among which the character of the photon momentum and also the Cherenkov effect stand out as typical examples. As noted before, the theory becomes nondispersive in the present formulation. Physically, this is a consequence of the external magnetic field ${\bf B}_0$ being treated as a static one. More general theories, where the magnetic field $\bf B$ (as well as an electric field $\bf E$) is time-dependent, would imply dispersive properties also in the magnetically-generated refractive index $n$.

Concerning the implication of the present work on Lorentz invariance violation (LIV), we would like to mention the following:

1. As we have shown, taking as an example the average magnetic field through which the gamma-ray burst has traveled through the cosmological distances to be $10^6$ Tesla, then they delay compared with the neutrinos to reach the Earth by a time in the range of $\tau = 8.7 \times (10^2 - 10^{18})$ seconds. Although a value as $10^6$ Tesla may look high, nevertheless it is well below the Schwinger critical limit. Such a delay is well compatible with the delays observed by ANTARES neutrino telescope \cite{A} or by the IceCube data \cite{B}.  

The observed time delays had originally been taken as an indication for possible LIV, interpreting the reduction in the speed of light to be the breaking of Special Relativity at high energies. Notice,  however, that in the interpretation of those observations, it has been assumed that the GRB and neutrino bursts have been emitted simultaneously, though it is even not known whether the GRB and the neutrino bursts have come from the same source or not.

Therefore, the results presented in this Letter put doubt on the claim that the GRB delay
compared with the neutrino bursts is a sign of LIV, since the delay can well be accounted
for by the standard physics, namely by the dispersion of light in a strong magnetic field found in the interstellar space due
to the existence of magnetars, as shown in the present work.

2. In the previous works \cite{chaichian20} and \cite{brevik21}, the dispersion of light and its delay were carefully studied by considering the effects of different media, electron-positron, CMB and the axion plasmas,  existing within the cosmic distances. The results in \cite{chaichian20} and \cite{brevik21} showed that, while the time  delay  of GRB by  the CMB medium was the largest compared with the other two media, still it was by far too small to account for the delays of GRB observed in \cite{A} and \cite{B}.

3. According to the dispersion relation in matter-filled
media studied in \cite{chaichian20} and \cite{brevik21}, as well as in any dispersive medium,  the more energetic the GRB, the less the  delay -
in other words, the more energetic GRB reach us earlier than the less energetic ones. Contrary to that,  the strong magnetic field considered in the present Letter produces out of the quantum vacuum a nondispersive medium, which does not depend on the energy of light and thus the refraction index depends only on the value of the external magnetic field.

4. Irrespective of the experimental observations of delay in GRB and also inspired by the results such as in \cite{A} and \cite{B}, there has been a remarkable activity, both on the theoretical and phenomenological possible occurrence of LIV at high energies. As a noncomprehesive list of references we mention \cite{E,F,G,  P,R,Z', Z'', Z'''}, and references therein.

We notice that in all the so-far proposed theoretical models, such as in a stringy space-time foam or in phenomenological models with some deformation of the usual dispersion relation $E^2={\bf p}^2$, and as well in the interpretation of the observations as arguments for LIV, either the space is not taken to be vacuum, but filled with some matter (i.e. the induced LIV), or the assumption of a simultaneous emission of particles with different energies in the gamma-ray bursts is used: For instance, typically in the LIV approach motivated by quantum gravity effects (which assumes the presence of gravity), the dispersion relation contains higher-powers in energy E, such as:
\begin{equation}\label{LIV_disp}
E^2\left[ 1+\xi\frac{E}{E_{QG}}+O\left(\frac{E^2}{E_{QG}^2}\right)
\right]={\bf p}^2c^2+m^2c^4,
\end{equation}
where $E_{QG}$ is an effective energy scale for quantum gravity,
commonly taken to be of the order $10^{16}\ \mathrm{GeV}$, and $\xi$ is an
arbitrary parameter.
Then, with  a modified  dispersion relation such as \eqref{LIV_disp}, in the  leading order in $E/E_{QG}$,  the group velocity becomes:
\begin{equation}
v_g= c\left( 1-\xi \frac{E}{E_{QG}}\right). \label{LIV}
\end{equation}
Other phenomenological models used for LIV based on the quantum deformation of the Lorentz or Poincar\'e groups lead to the same effect as \eqref{LIV} does, i.e. increasing the gamma-ray  energy makes the velocity smaller.

 According to \eqref{LIV}, the more energetic the GRB, the later they reach the Earth. At the same time the analyses of GRB time delay as a function of their energy also show that  the higher energy gamma-rays arrive later. Now, if one assumes that the GRB with all different energies have been emitted by a single source simultaneously, then of course one {\it can} attribute such a observation to a LIV, which is also in accord with \eqref{LIV}.

 But could one justify such an assumption that all the GRB with  different energies have been emitted  by a single source {\it simultaneously}?

We do not have any clue about the dynamics of bursts and the emission of particles from their sources, as they are truly complicated. But what we do know from the other sources, such as from the 11-Year Solar Cycle and also from the atomic explosions, is that the lower energy particles are emitted first, while the higher energy ones later. The latter phenomenon can be naturally explained when one acknowledges that the temperatures in such explosive sources are much higher in their centre and also that the diameters of such sources are usually huge.  This puts doubt on  the assumption of simultaneous emission, and consequently on  claiming the observation  of  Lorentz invariance violation  based on the GRB time delays.

Our conclusion is that there is no evidence for the Lorentz invariance violation, i.e. for
the basic assumption of the Special Relativity in the vacuum\footnote{We should keep in mind that, if  the  Lorentz invariance of the Special Relativity is violated, then it automatically implies that  the diffeormorphism invariance
of the General Relativity is also broken. Therefore, the Eintein gravitational theory and its modifications should be accordingly changed too, and only the latter theories with diffeormorphism symmetry broken should be used in all the applications.} -- all the data can be explained
by the standard physics, unless one assumes the existence of some kind of matters in the space
and invokes the use of an unjustified assumption.

\section*{Acknowledgements}
We are most grateful to Efrain Ferrer, Kimball Milton,
Viatcheslav Mukhanov, Olivier Piguet, Adam Schwimmer, Jenny Wagner and in particular to Felix Karbstein, Yuri Obukhov and Misao Sasaki
 for several illuminating discussions and suggestions.

\end{document}